\documentclass[%
 reprint,
 amsmath,amssymb,
 aps,
]{revtex4-1}

\usepackage{graphicx}
\usepackage{dcolumn}
\usepackage{bm}

\newcommand{\transition}[2]{$\mid\!\!#1\rangle\rightarrow\mid\!\!#2\rangle$}
\newcommand{\ket}[1]{$\mid\!\!#1\rangle$}

\begin{document}

\preprint{APS/123-QED}

\title{Observation of atomic localization using \\Electromagnetically Induced Transparency}

\author{N. A. Proite}
\author{Z. J. Simmons}%
\author{D. D. Yavuz}

\affiliation{%
Department of Physics \\
University of Wisconsin - Madison, 1150 University Ave.~Madison, WI 53706, USA
}%

\date{\today}

\begin{abstract}

We present a proof-of-principle experiment in which the population of an atomic level is spatially localized using the technique of  electromagnetically-induced transparency (EIT). The key idea is to utilize the sensitive dependence of the dark state of EIT on the intensity of the coupling laser beam. By using a sinusoidal intensity variation (standing-wave), we demonstrate that the population of a specific hyperfine level can be localized much tighter than the spatial period. 

\end{abstract}

\maketitle


It is well-known that traditional optical techniques cannot resolve or write features smaller than half the wavelength of light. This barrier, known as the diffraction limit, has important implications for a variety of scientific research areas including biological microscopy and quantum computation. As an example, in a neutral-atom quantum computing architecture, the diffraction limit prohibits high-fidelity manipulation of individual atoms if they are separated by less than the wavelength of light. Recently, Agarwal and others \cite{agarwal,gorshkov,yavuz} have proposed to use the dark state of electromagnetically induced transparency (EIT) \cite{harris,scully} to address atoms at potentially nanometer spatial scales. This technique relies on the sensitive dependence of the dark state to the intensities of the driving probe and coupling laser beams. If a standing-wave coupling laser is used, the population of the excited Raman level can be very tightly localized near the intensity nodes, allowing for sub-wavelength control. In this letter, we present a proof-of-principle experiment that demonstrates the key ideas of this approach. By using ultracold Rubidium (Rb) atoms in a magneto-optical trap (MOT) and pulsed coherent transfer, we demonstrate atomic localization to spots much smaller than the spatial period of the coupling-laser intensity profile. Although due to imaging limitations we have used a large spatial period in this work ($\approx 600~\mu$m), our results will likely scale to the sub-wavelength regime in the future. 

Before proceeding, we cite important prior work leading up to this experiment.  In their pioneering work, Thomas and colleagues have suggested and experimentally demonstrated sub-wavelength position localization of atoms using spatially varying energy shifts \cite{thomas,stokes,gardner}. Zubairy and coworkers have discussed atom localization using resonance fluorescence and phase and amplitude control of the absorption spectrum \cite{kiffner,macovei,kapale_zubairy}. Knight and colleagues discussed localization via quantum interference at the probability amplitude of the excited electronic state \cite{paspalakis}. Li et.~al.~have experimentally demonstrated probe narrowing beyond the diffraction limit using a spatially-varying coupling laser profile in a vapor cell \cite{li}.  There has also been remarkable progress in utilizing the position dependent stimulated emission to achieve nanoscale resolution \cite{hell,maurer}. This last approach, also known as stimulated-emission depletion microscopy, is now a widely used technique in biological imaging. We note that our approach of using the dark state for atomic localization has the following key advantages: 1) For the ideal case of sufficiently slowly varying driving laser pulses, the dark-state technique has no population at the excited electronic state. As a result, the atomic localization can, in principle, be achieved without suffering from the detrimental effects of spontaneous emission. This is especially important for quantum computing applications \cite{gorshkov} where coherent manipulation with little decoherence is essential. 2) The dark state can be prepared adiabatically by using a counter-intuitive pulse sequence. As a result, as discussed in detail in Ref.~\cite{yavuz}, the scheme is insensitive to many experimental fluctuations such as the intensity and the timing jitter of the driving laser pulses. 3) Since the scheme is coherent, localization can be achieved at faster time-scales at the expense of requiring more intense laser beams. Although in this work we use $\approx 100$~ns-long laser pulses, dark-state-based localization can easily be achieved at sub-ns time-scales by using more powerful laser beams. 

\begin{figure}[tb]
\includegraphics[width=8.3cm]{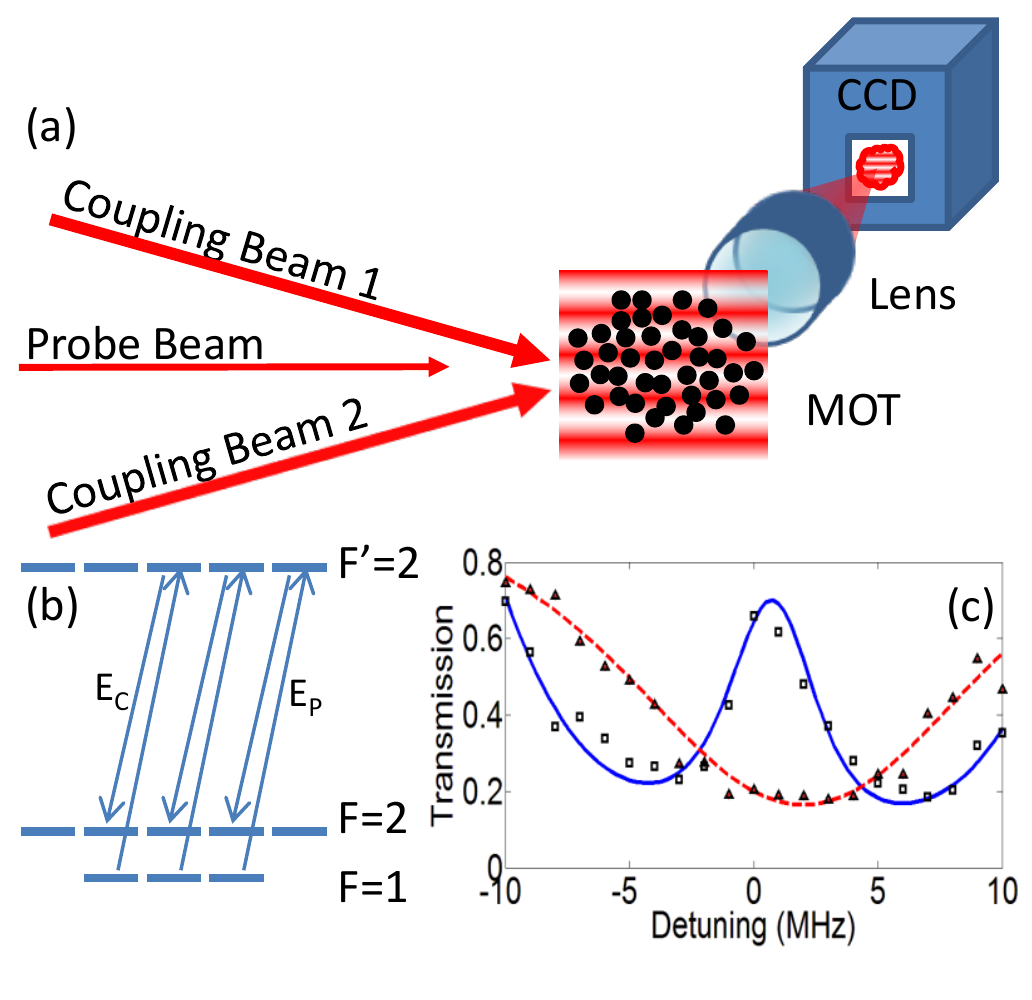}
\caption{\label{setup} (Color online) (a) The schematic of our experiment. The experiment is performed in a magneto-optical trap (MOT) of ultracold $^{87}$Rb atoms. With atoms starting in the \ket{F=1} hyperfine level, we drive the atoms to the dark state with a probe laser beam and a spatially-varying coupling laser beam. The spatial profile for the coupling laser is obtained by combining two identical beams at the MOT at an angle of 3 milliradians producing a vertical standing wave. The atomic localization is measured by taking fluorescence images with the CCD. (b) The relevant energy level diagram, with probe laser E$_P$ and coupling laser E$_C$. The experiment works with three  parallel $m_F$ channels. (c) Transmission of a weak probe beam ($\approx$ 100~nW power) through the cloud as a function of frequency with (solid blue line) and without (dashed red line) the coupling laser beam. We perform this measurement with Coupling Beam 1 at an intensity of 120 mW/cm$^2$ and do not use the standing-wave.  The on-resonance transmission is about 70\% demonstrating reasonably good EIT.}
\end{figure}

We next discuss the details of our experiment which can be viewed as a proof-of-principle demonstration of the suggestion by Lukin and colleagues \cite{gorshkov}.  The experiment is performed inside a 14-port, stainless-steel ultrahigh vacuum chamber which is kept at a base pressure of about 10$^{-9}$ torr. To construct the $^{87}$Rb MOT, we use three counter-propagating beam pairs that are locked to the cycling transition, each with a beam power of 100~mW and a beam size of 3~cm. The MOT lasers are obtained from an external cavity diode laser whose output is amplified with a tapered amplifier. We typically trap about 1~billion $^{87}$Rb atoms at a temperature of 150~$\mu$K. The EIT beams are derived from a separate master diode laser which is saturated-absorption locked to the appropriate transition.  The coupling laser beam is shifted by 6.8 GHz using high-frequency acousto-optic modulators and is amplified with a tapered amplifier \cite{unks}.  As shown in Fig.~\ref{setup}, the probe and the coupling lasers are resonant with \transition{F=1}{F'=2} and \transition{F=2}{F'=2} transitions of the D$_2$ line, near a wavelength of 780 nm.  The beams have the same circular polarization and the experiment works in three parallel $m_F$ channels \cite{braje}. The coupling laser is split into two beams, which then reconverge at the MOT at an angle of 3 milliradians to form a vertical standing wave with a spatial period of $\Lambda=600~\mu$m.  We probe the localization by level-dependent fluorescence of the atomic cloud.  The fluorescence signal is collected with a 2-inch achromatic-doublet outside of the vacuum chamber and is recorded with an electron-multiplying CCD camera.

Before proceeding further, we present a brief discussion of population localization using the dark state. Atoms distributed throughout the MOT will see different coupling laser intensities, based on where they are in the standing wave. Ignoring the complications due to parallel channels, the dark state of the atoms is given by \cite{harris,gorshkov,yavuz}:
\begin{eqnarray}
\mid\!\text{dark(x)}\rangle=\frac{\Omega_C(x)\mid\!\!F=1\rangle -\Omega_P\mid\!\!F=2\rangle}{\sqrt{\Omega_{C}(x)^2 + \Omega_P^2}}, 
\end{eqnarray}
where $\Omega_P$ and $\Omega_C$ are the Rabi frequencies of the respective beams. Here, for simplicity, we assume the probe beam to be uniform.  The atoms can be prepared in the dark state of Eq.~(1) by using the well-known counter-intuitive pulse sequence with coupling laser turning on before the probe laser beam. Once the laser beams are turned on, they can be turned-off simultaneously preserving the ratio of the Rabi frequencies \cite{yavuz}. As a result, even after the laser pulses are turned-off, the atomic system is left in the state as determined by the probe and coupling laser Rabi frequencies at the temporal-peak of the pulses. Through this preparation, atoms will populate \ket{F=2} with a probability of $|\langle F=2|\text{dark(x)}\rangle|^2=\Omega_P^2/\left[\Omega_C(x)^2+\Omega_P^2\right]$.  Due to the sensitive dependence to the coupling beam intensity, atoms located near a coupling field zero-crossing (intensity node) coherently transfer to \ket{F=2} with high probability.  If we assume that $\Omega_C (x)$ is linear near a zero-crossing, then we expect the probability $|\langle F=2|\text{dark(x)}\rangle|^2$ to be maximum at the coupling intensity node, and have an approximate spatial width of $\sim\!\!\Lambda \cdot\Omega_P/\Omega_{C0}$ where $\Omega_{C0}$ is the peak coupling laser Rabi frequency \cite{gorshkov,yavuz}. As a result, with the probe laser intensity fixed, the population of level \ket{F=2} becomes more and more localized with increasing coupling beam power.  

\begin{figure}[tb]
\includegraphics[width=8.3cm]{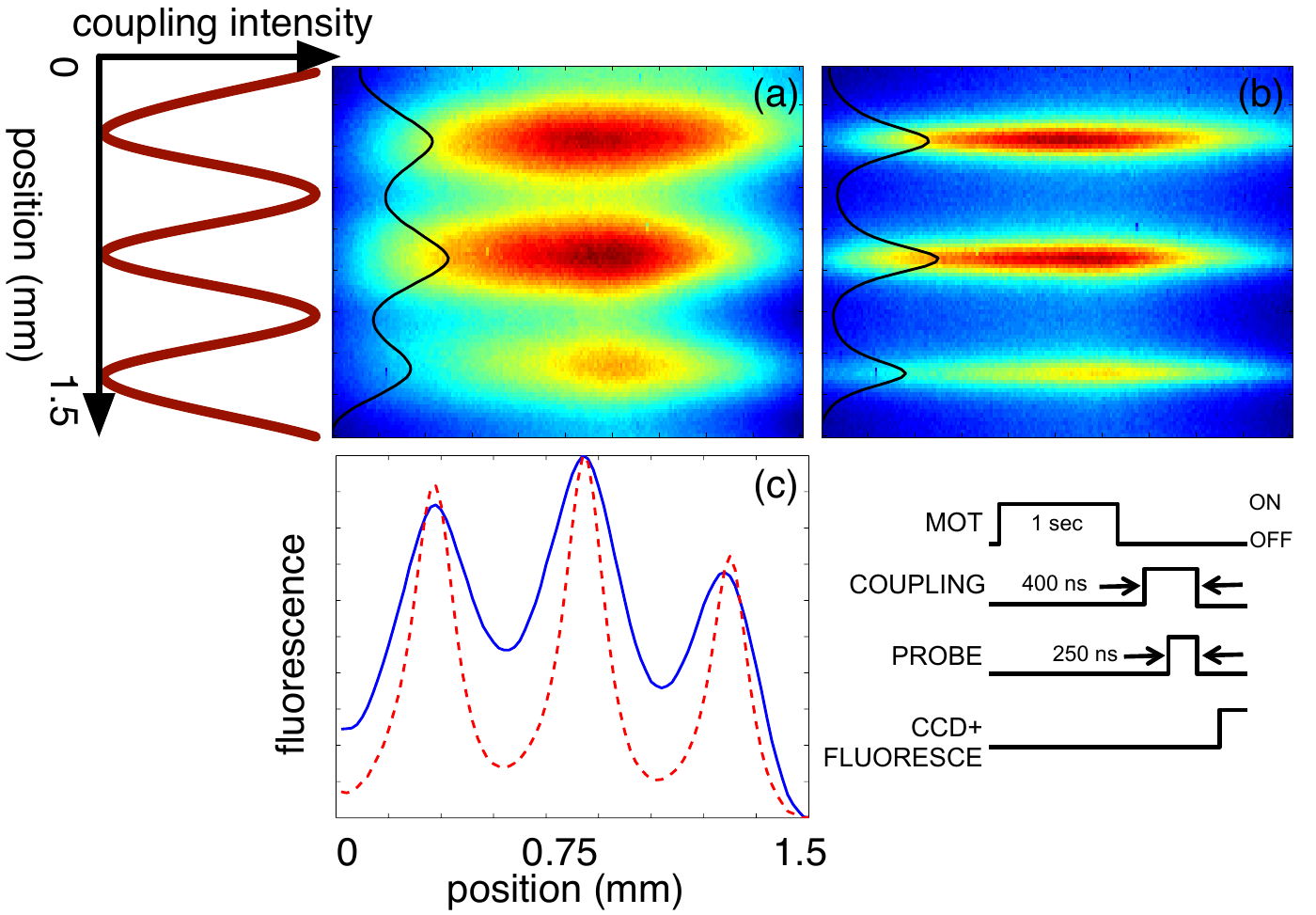}
\caption{\label{ccd} (Color online) Fluorescence images of the atomic cloud for (a) $I_{C0}\simeq 22\times I_P$ and (b) $I_{C0}\simeq 418\times I_P$. The images are obtained by fluorescing the  \ket{F=2} level via the cycling transition after the EIT beams are turned-off. The fringes are confined to the intensity nodes of the coupling beam and become more localized as the intensity of the coupling laser increases. (c) shows horizontally-averaged line profiles of each fluorescence image for more direct comparison. The solid line is for part (a) and the dashed line is for part (b). The lower right diagram shows the experimental timing cycle.}
\end{figure}

The experimental timing cycle is shown in Fig.~\ref{ccd}. We begin the experiment by loading the MOT for one second and then turn off the MOT magnetic field gradient 50 ms prior to the EIT beams to reduce Zeeman splitting of the magnetic sublevels.  All atoms are then initialized to \ket{F=1} by turning off the hyperfine repumping laser for the MOT.  We drive the atoms to the dark state by using a 400 ns-long coupling laser and a 250 ns-long probe laser beam.  After the EIT beams are turned-off, we probe the population of \ket{F=2} by fluorescing the atoms for 40~$\mu$s via the cycling transition (\transition{F=2}{F'=3}). Due to sufficiently low atomic temperature, the motion of the atoms during fluorescence is negligible. In Fig.~\ref{ccd} we present two fluorescence images that show localization of the \ket{F=2} population as the coupling laser intensity is increased.  Fig.~\ref{ccd}(a) illustrates a case where we use a relatively weak coupling beam, where $I_{C0}\simeq 22\times I_P$ ($I_{C0}$ is the peak coupling intensity and $I_P$ is the probe intensity). The fringes align with the nodes of the coupling beam intensity and have wide profiles in the vertical dimension.   In Fig.~\ref{ccd}(b), we use a nearly 20 times more intense coupling laser beam such that $I_{C0}\simeq 418 \times I_{P}$. We observe the fringes to be vertically much more tightly confined to the coupling beam nodes. Both pictures use a probe intensity of 3.9 mW/cm$^2$, and each picture is an average of 100 shots.  Fig.~\ref{ccd}(c) shows horizontally-averaged line profiles of each fluorescence image for more direct comparison.  

\begin{figure}[tb]
\includegraphics[width=8.3cm]{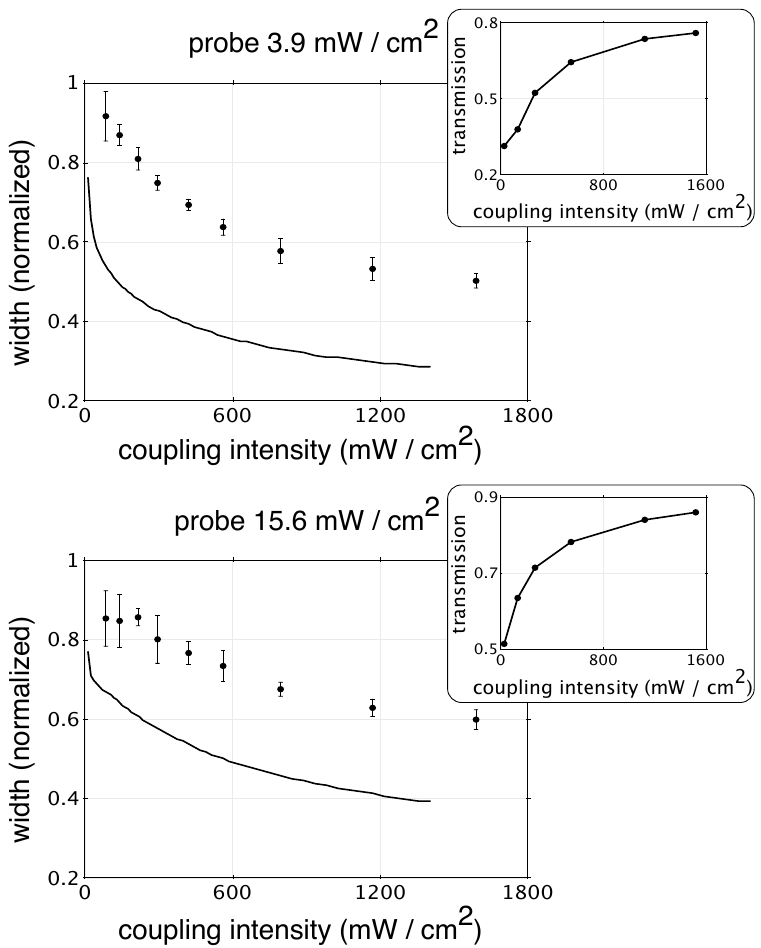}
\caption{\label{data} The width of the fringes as a function of peak coupling beam intensity for $I_P=3.9$~mW/cm$^2$ and $I_P=15.6$~mW/cm$^2$. The vertical scale is normalized such that one unit corresponds to a fringe width that equals half the spatial period (a sine wave). The population of  \ket{F=2} becomes more tightly localized to the standing wave nodes with increasing coupling laser beam intensity. The data points are experimental observations and the solid lines are the result of a numerical simulation. We attribute the discrepancy between experiment and theory to various imperfections such as the mechanical jitter of the standing-wave pattern. See text for details. The insets show the integrated probe transmission through the cloud as the intensity of the coupling beam is increased. We observe increased transmission with increasing coupling intensity, demonstrating the presence EIT.}
\end{figure}

Figure \ref{data} shows the normalized full-width-half-maximum (FWHM) of the fringes as the coupling beam intensity is varied for two values of probe laser intensity $I_P=3.9$~mW/cm$^2$ and $I_P=15.6$~mW/cm$^2$. Each data point is an average of 100 images and the error bars show the standard deviation of each set. For $I_P=3.9$~mW/cm$^2$, we observe the population of level \ket{F=2} to localize by about a  factor of two as the coupling beam intensity is increased. The solid lines in Fig.~\ref{data} are the results of numerical calculations without any adjustable parameters (i.~e.~each parameter that goes into the simulations are experimentally measured). Here, we include all relevant magnetic sub-levels and numerically solve the time-domain density matrix equations for the conditions of our experiment. We have experimentally measured the standing-wave interference of the coupling laser beam to be slightly imperfect with intensity contrast of 98\%. This imperfection is included in our numerical calculations. The disagreement between theory and experiment is likely a result of 1) mechanical and interferometric fluctuations of the standing-wave intensity profile of the coupling laser beam, and 2) the Zeeman shift of the magnetic-sublevels due to a residual background magnetic field.
 
\begin{figure}[tb]
\includegraphics[width=8.3cm]{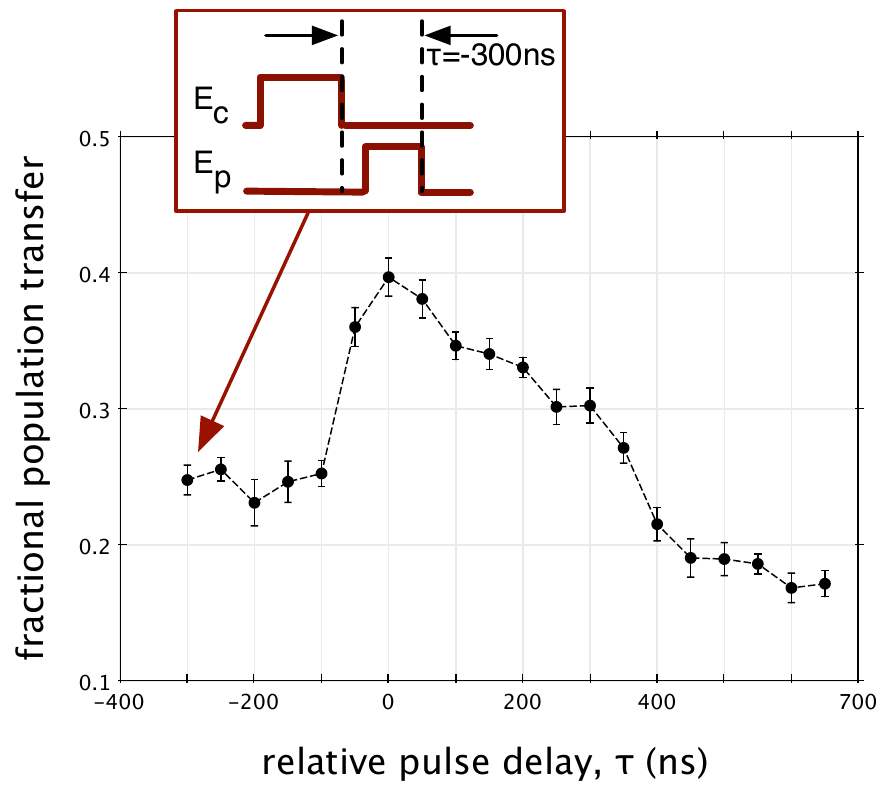}
\caption{\label{stirap} (Color online) Demonstration of stimulated Raman adiabatic passage (STIRAP).  For this test, we initialize the atoms to level \ket{F=1} and probe the population of \ket{F=2} as the overlap between the EIT pulses is varied.  We observe maximal population transfer to \ket{F=2} when the coupling and probe pulses completely overlap with coupling laser turning-on and turning-off before the probe laser beam.  }
\end{figure}

We next discuss the coherent nature of population localization. The insets in Fig.~\ref{data} show the integrated probe transmission through the atomic cloud as the coupling beam intensity is increased. We see better probe transmission with increased coupling beam intensity, demonstrating EIT for the exact conditions of each localization experiment. Furthermore, we have the ability to probe excited state fluorescence during the EIT process by collecting scattered photons for the duration of the coupling laser beam. We observe a reduction in the excited state fluorescence as the coupling laser intensity is increased, complementing the probe transmission data of the insets of Fig.~\ref{data}. We also observe a strong increase in the excited state fluorescence when the coupling laser beam is turned-off (probe laser propagating alone through the cloud). This further confirms that the atoms are driven to a dark state with a small population at the excited electronic level.  

To further test the coherent nature of the population transfer, we have also performed a stimulated Raman adiabatic passage (STIRAP) experiment \cite{bergmann}. We measure the population transfer to \ket{F=2} at the intensity peaks of the coupling laser using a pulse sequence similar to above, but by changing the relative temporal overlap between the EIT beams. Noting Fig.~4, as expected, the maximum transfer to \ket{F=2} occurs when the probe and coupling pulses overlap, with coupling laser turning-on and turning-off before the probe laser beam. We observe a 20\% increase in population transfer when the two pulses overlap, consistent within a factor of two of our density-matrix calculations. Near the intensity nodes of the coupling laser, we observe approximately 10\% increase in population transfer when the pulses overlap (not shown in Fig.~\ref{stirap}).  As mentioned earlier, there is a coupling beam intensity offset of 2\% of the peak at the nodes due to an imperfect interference profile.  To increase contrast, the STIRAP experiments of Fig.~4 use beams that are 12~MHz detuned to the blue of the excited state. The intensities of the two beams are I$_{C0}\approx \text{I}_P$ =  130 mW/cm$^{2}$.

To summarize, we have demonstrated localization of level population using EIT. As mentioned before, because our imaging system cannot resolve sub-wavelength spatial scales, we have performed this experiment with a small-angle between the two coupling-laser beams and therefore with a large spatial period of the standing-wave interference pattern. Future work will include extending this technique to the sub-wavelength regime and possibly demonstrate nanometer scale localization and addressing of neutral atoms. Furthermore, by using more powerful laser beams, we aim to explore atomic localization at much faster time-scales. If successful, the ability to address atoms at sub-ns time-scales with sub-wavelength resolution may provide a powerful tool for many challenging problems including initialization and addressability of a neutral-atom quantum register \cite{urban,gaetan}. 

We thank J.~P.~Sheehan for assistance with the experiment.  This work was supported by the Air Force Office of Scientific Research (AFOSR).

\providecommand{\noopsort}[1]{}\providecommand{\singleletter}[1]{#1}%

\end{document}